\documentclass[prl,twocolumn]{revtex4}
\usepackage{amsmath}
\usepackage{epsf}

\begin{document}

\title{On the Use of Finite-Size Scaling to Measure Spin-Glass Exponents} 
\author{A. C. Carter}
\author{A. J. Bray}
\author{M. A. Moore}
\affiliation{Department of Physics \& Astronomy, The University of
Manchester, Manchester M13 9PL, UK}
\date{\today}

\pacs{75.10.Jm, 75.50.Lk}
\begin{abstract}

Finite-size  scaling  (FSS)  is  a standard  technique  for  measuring
scaling exponents in spin glasses.  Here we present a critique of this
approach,  emphasizing the  need for  all  length scales  to be  large
compared  to  microscopic  scales.  In  particular we  show  that  the
replacement,  in  FSS  analyses,  of  the correlation  length  by  its
asymptotic scaling form can  lead to apparently good scaling collapses
with the wrong values of the scaling exponents.

\end{abstract}

\maketitle

This  paper is  a critique  of  finite-size scaling  (FSS) methods  as
applied   to  spin   glasses.   We   focus  primarily   on   one-  and
two-dimensional Ising  spin glasses,  which order only  at temperature
$T=0$, though  the underlying  ideas are quite  general. The  model is
defined by  the Hamiltonian  $H= - \sum_{<i,j>}J_{ij}S_iS_j  - h\sum_i
S_i$, where $h$ is an external  field which will be zero unless stated
otherwise.

In  dimension  $d=2$,  a  range  of different  methods  for  measuring
exponents has led  to widely differing quoted values,  sometimes by as
much    as     a    factor     two,    for    the     same    exponent
\cite{BM84,McMillan84,HM85,BhattY88,Kawashima92,Rieger96,HuseKo97,Ritort}. 
Despite years  of numerical study,  these   discrepancies have  never been
satisfactorily  resolved.  Here  we present  analytical  and numerical
results in dimension $d=1$, for  which exact analytical values for the
scaling exponents are  known. By applying standard FSS  methods to the
data, we  find large discrepancies between  the numerically determined
exponents  and  the exact  ones.   We  show  that these  discrepancies
disappear if the data are plotted in a different  way, using the exact
correlation length in the FSS  analysis instead of its leading scaling
form. We conclude  that the main cause of  the discrepancies is  large
corrections to  scaling in the expression for  the correlation length,
rather than corrections to FSS itself.

For the critical behavior as $T  \to 0$, there is only one independent
exponent \cite{Binder}, except perhaps  when the ground state has
a non-trivial degeneracy. The exponent on which we focus our attention
is the `stiffness exponent' $\theta$, which describes the dependence on
length  scale, $l$,  of the  energy, $E$,  of an  excitation  from the
ground state: $E \sim l^\theta$ \cite{BM84,McMillan84,Heidelberg,FH86}. 
For $d \le 2$, $\theta$ is negative and large excitations are easily 
created by thermal fluctuations, destroying the
ground-state order.  Setting $E \sim T$ gives  a characteristic length
scale $\xi  \sim T^{-\nu}$,  with $\nu =  -1/\theta$, above  which the
ground state is unstable against thermal fluctuations, i.e.\ $\xi$ can
be identified with the correlation length \cite{BM84,Heidelberg}.

The exponent $\theta$ describes the dependence on $l$ of the energy of
a  `droplet' of reversed  spins, of  linear size  $l$.  The  energy is
associated with the boundary of the droplet. A measurement of $\theta$
can be  made through the numerical  study of domain  wall energies, in
which a  wall, or `interface', is  imposed on a  finite-size system by
choice of boundary conditions \cite{BM84}. One finds $\theta \approx
-0.28$  in  $d=2$. It  was  recently shown  that  this  result can  be
obtained in  a way  which is explictly  independent of  the particular
boundary conditions  used to impose  the interface \cite{CartEtAl02}.
The scaling  result $\nu =  -1/\theta$ then predicts $\nu \approx 3.6$
for the  correlation length exponent. 

There has been some debate as to whether the exponent $\theta$ (sometimes 
called $\theta_{DW}$) extracted from domain wall energies is identical to 
the exponent obtained by studying directly the size dependence of droplet 
energies. Droplet excitations can be created from the ground state by 
reversing a central spin holding the boundary spins fixed, where the scale 
$l$ is the linear size of the system \cite{Kawashima99}. This approach gives 
$\theta \approx -0.42$, though the analysis has been criticised by Middleton 
who argues that only droplets whose area exceeds some fraction of $l^2$
should be included \cite{Middleton}. Hartmann and Moore \cite{HartMoore} have 
shown how the apparent difference between the two values of $\theta$ can 
be explained by invoking a correction to scaling (a subdominant term of 
order $l^{-\omega}$, with $\omega > - \theta$, in the expression for 
the droplet energy), and that the corresponding value of $\theta$ is 
consistent with that obtained from domain wall studies. Such corrections 
to scaling may also account for the value $\theta \approx -0.46$ obtained 
by Picco et al.\ in their recent study of droplet energies \cite{Ritort}. 

A `direct' measurement of the exponent $\nu$,  that describes the 
divergence of the correlation length, can be made by studying, for example,
the spin-glass susceptibility
\begin{equation}
\chi_{SG} = L^{-d} \sum_{i,j} \overline{\langle S_iS_j \rangle^2}\ ,
\label{chi}
\end{equation}
where the angular brackets represent a thermal average, the overbar is
a disorder average, and $L$ is the linear size of the system. At $T=0$
every  term in  the  sum  is unity  and  $\chi_{SG}=L^d$.  For  $T>0$,
$\overline{\langle S_iS_j  \rangle^2} =  f(r/\xi)$, where $f(x)$  is a
scaling function  with $f(0)=1$, $f(x)  \sim \exp(-x)$ for  large $x$,
giving \cite{Binder} $\chi_{SG} \sim \xi^d \sim T^{-\gamma}$ with
$\gamma=d\nu$.   FSS predicts $\chi_{SG}  = L^d  F(L/\xi)$ for  $L \to
\infty$, $\xi \to \infty$ with $L/\xi$ fixed but arbitrary. Using $\xi
\sim T^{-\nu} = T^{1/\theta}$, one can  determine $\theta$ by plotting 
$L^{-d}\chi_{SG}$
against $TL^{-\theta}$, and  choosing $\theta$ to give the  best data 
collapse. Using  this  method, Kawashima  et al.\ \cite{Kawashima92}  
find  $\nu \approx 2.0$, i.e.\ $\theta \approx -0.5$, which differs  
significantly from the value $\simeq -0.28$ inferred from the domain-wall 
studies.

A second method that has been  used to extract $\theta$ numerically is 
to measure the  magnetization   per   spin,   $m(h)   =   L^{-d}\sum_i
\overline{\langle S_i \rangle}$, induced by a small magnetic field $h$
at $T=0$ 
\cite{Rieger96,Baharona94,McMillan84a,Reger84,Kawashima92a}. Since  
in zero field we have $m(0)  \sim L^{-d/2}$, a simple scaling argument  
for small $h$ gives $m(h)  = L^{-d/2} g[h(L)/J(L)]$,
where $h(L)  \sim hL^{d/2}$  is the effective  field at scale  $L$ and
$J(L) \sim  L^\theta$ is the  effective coupling at this  scale. Hence
$m(h)  = L^{-d/2}M(L^{d/2-\theta}h)$. In  the thermodynamic  limit $m$
should  be   become  independent  of   $L$,  which  implies   $m  \sim
h^{1/\delta}$ with  $\delta = 1-2\theta/d \approx 1.28$  for $d=2$. An
alternative   way   of   writing    the   FSS   form   is   
\begin{equation}
m_L(h) = L^{-d/2}G(L^{d/2}h^{1/\delta})\ , 
\label{m}
\end{equation}
with $G(0+)  = {\rm const}$ and $G(x)
\sim x$ for $x \to \infty$.  With this method, Rieger et al.\ 
\cite{Rieger96} obtained $\delta \approx  1.48$, which  differs 
significantly from  the scaling prediction $\approx 1.28$ and, naively, 
predicts that $\theta \approx -0.48$ instead of the value $\approx -0.28$ 
obtained from domain-wall studies. 

To summarise, different ways of measuring $\theta$ give different, and 
seemingly incompatible, results. The value $\nu \approx 2$ obtained in 
\cite{Kawashima92} is equivalent to $\theta \approx -0.5$, which is 
similar to the value $\theta \approx -0.48$ inferred from the $m(h)$ 
data of \cite{Rieger96}, and to the values $\approx -0.42$ and $\approx 
-0.46$ obtained from studying droplet excitation in \cite{Kawashima99} 
and \cite{Ritort} respectively, all of which differ from the value 
$\approx -0.28$ obtained from domain-wall studies \cite{BM84,McMillan84, 
CartEtAl02}. 

In an attempt to understand these differences, we have carried out 
analytical and numerical studies in space dimension $d=1$, for which 
the corresponding exponent values are known exactly. We mimic 
the two-dimensional studies of Kawashima et al \cite{Kawashima92} and 
Rieger at al \cite{Rieger96}, and look at the $T$-dependence of 
$\chi_{SG}$ at $h=0$ and the $h$-dependence of $m$ at $T=0$ respectively. 
We conclude that these that these quantities are affected by  
corrections to scaling so large that it is essentially impossible to 
extract the correct exponent values from system sizes that are accessible 
in $d=2$. 

In the remainder of the paper, therefore, we consider the  $d=1$ Ising 
spin glass in two situations:  (i) $T>0$ and $h=0$. We  calculate 
$\chi_{SG}$ and use the FSS form $\chi_{SG}  = Lf(TL^{-\theta})$ to 
determine $\theta$. We show that the exact value of $\theta$  gives a 
very poor data collapse for the system sizes studied, and that  a 
reasonable data collapse is obtained with a significantly  different 
value of $\theta$; (ii)  $T=0$, $h>0$. We
calculate  $m_L(h)$  and  use  the FSS  form  $m_L(h)=L^{1/2}g(L^{1/2}
h^{1/\delta})$ to determine $\delta$. Again, the exact $\delta$ gives a
poor collapse, and the best collapse is obtained with a very different
$\delta$. To facilitate comparison with  the $d=2$ data of Kawashima et 
al and of Rieger  et al.\  we choose,  for the  $h=0$ results,  a bond
distribution  $P(J)$ engineered  to give  $\theta =  -0.282$,  i.e.\ a
value equal to  that of the $d=2$ system, while for  the $h>0$ data we
choose the  distribution such  that $\delta =  1.282$ is equal  to the
value predicted in $d=2$ for  $\theta = -0.282$. These choices are are
imposed by using a bond  distribution of the form 
\begin{equation}
P_\alpha(J) \propto |J|^\alpha\exp(-J^2/2)\ .
\label{P}
\end{equation}  
For  any distribution satisfying  $P(J) \sim
|J|^\alpha$  for $J  \to 0$  it may  be shown \cite{Heidelberg} that
$\theta =  -1/(1+\alpha)$ for  $d=1$, so the  choice $\alpha  = 2.546$
gives  $\theta =  -0.282$,  while  $\alpha =  6.042$  gives $\delta  =
1.282$.

\begin{figure}
\begin{center}
\epsfxsize=80mm
\epsfbox{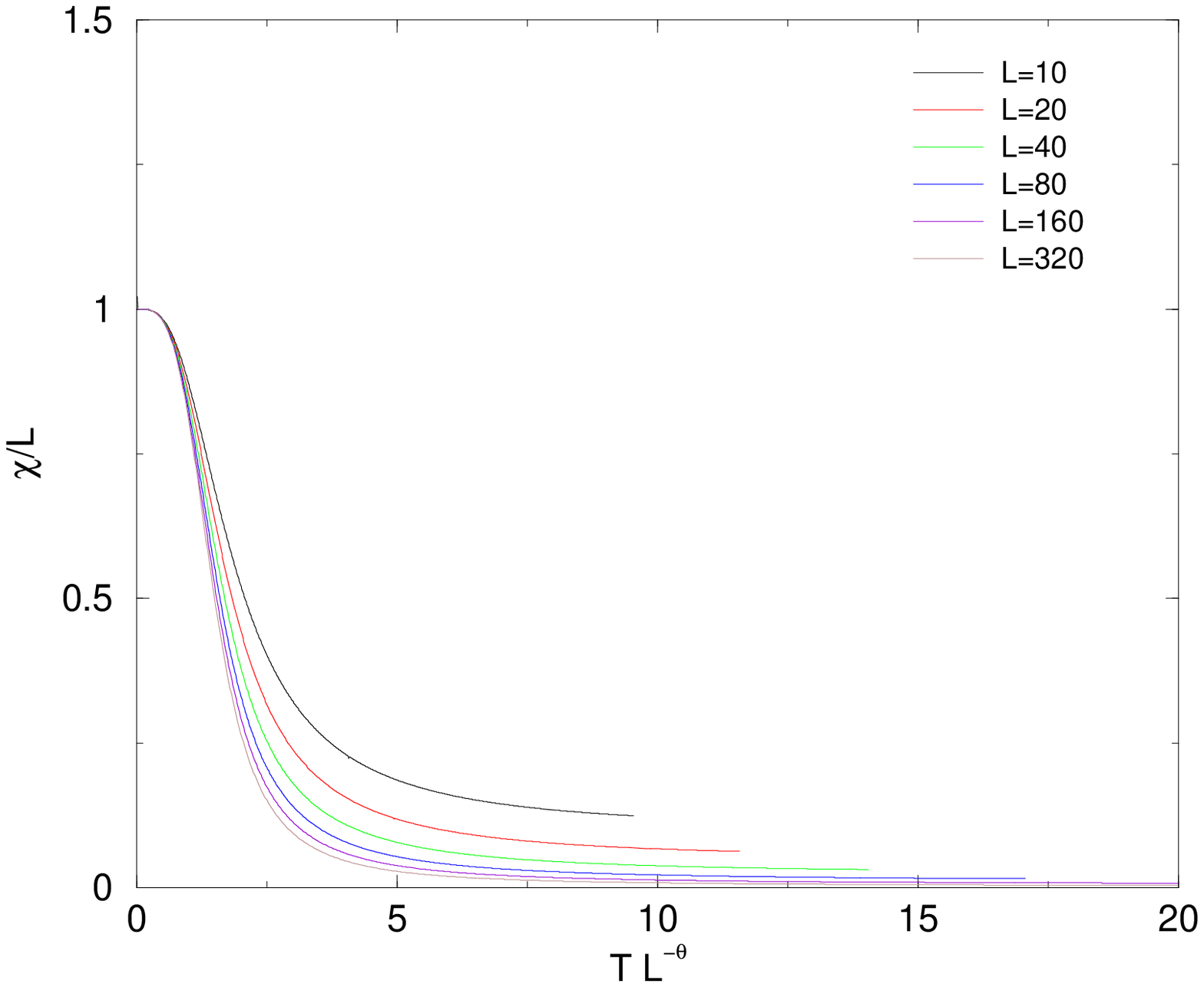}
\epsfxsize=80mm
\epsfbox{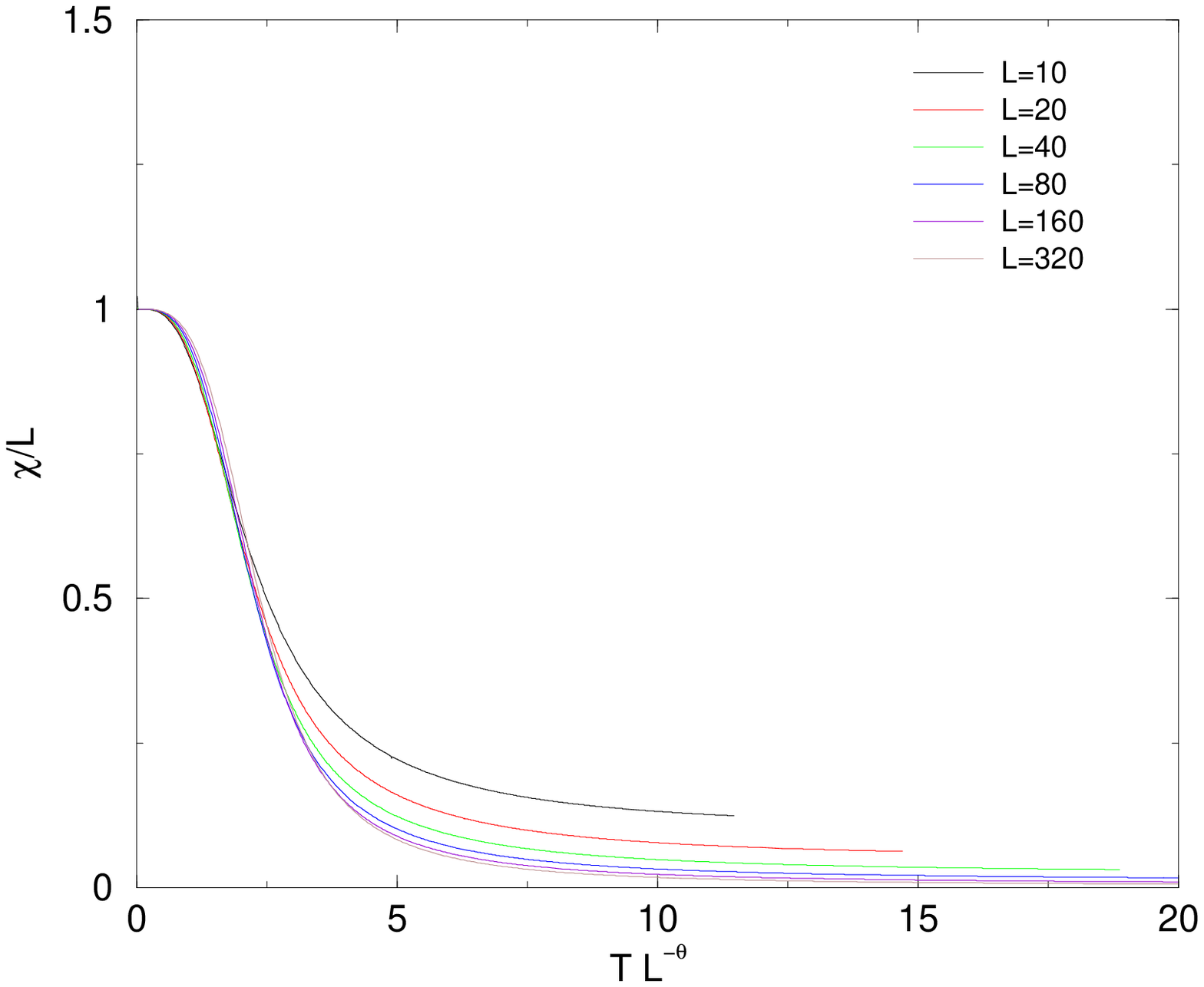}
\caption{Susceptibility scaling plot. From top to bottom: 
$\theta=-0.28$ (theoretical) and $\theta=-0.36$ (best fit by eye). 
\label{fig:Tcol}}
\end{center}
\end{figure}

\begin{figure}
\begin{center}
\epsfxsize=80mm
\epsfbox{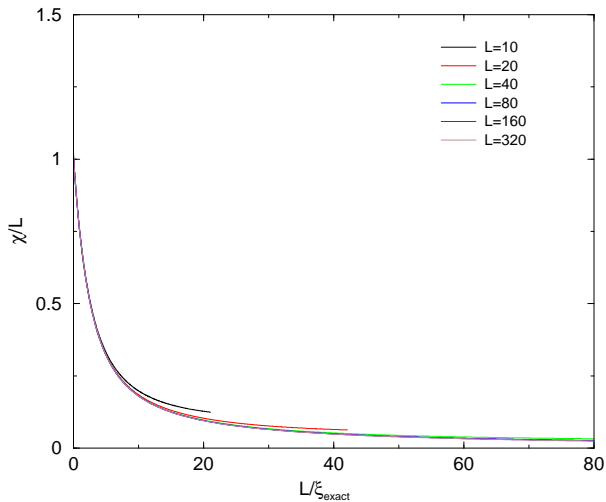}
\caption{Data collapse with scaling variable $L/\xi_{exact}$\label{fig:xx}}
\end{center}
\end{figure}

\smallskip
\noindent\underline{\bf (i) $T>0$, $h=0$} \\ The Ising Hamiltonian for
$d=1$  can  be  written  $H=-\sum_iJ_iS_iS_{i+1}$,  and  we  use  free
boundary conditions.  It is straightforward  to show that (for  $j \ge
i$) $\langle S_iS_j \rangle  = \prod_{r=i}^{j-1} \tanh(\beta J_i)$ and
\begin{equation}
\overline{\langle S_iS_j \rangle^2} = a^{|j-i|}\ ,
\end{equation}
where $a=\overline{\tanh^2(\beta J_r)}$ is independent of $r$. Finally
\begin{equation}
\chi_{SG}=\frac{1}{L}\sum_{i,j=1}^L\overline{\langle S_iS_j\rangle^2}  
  = \frac{1+a}{1-a} - \frac{2a(1-a^L)}{L(1-a)^2}.
\end{equation}
The quantity $a=\int dJ\,P_\alpha(J)\tanh^2(\beta J)$ may be evaluated
numerically  for any  temperature  $T$. The  resulting $\chi_{SG}$  is
plotted  in   Fig.\ 1, in  the  scaling  form   $\chi_{SG}/L$  against
$TL^{-\theta}$, for lattice sizes up  to $L=320$. In the upper figure,
we use the  exact value, $\theta = -0.282$, while in  the lower we use
$\theta=-0.36$, which  is our `best  fit by eye'. Comparing  these two
plots one observes  that (i) The exact value of  $\theta$ gives a good
collapse  only  at  small   values  of  the  scaling  variable,  where
$\chi_{SG}/L$ is close to its $T=0$ value of unity; (ii) The `best fit
by eye' is a much better fit over a large part of the plot, especially 
for the larger systems. Small but systematic departures from perfect 
scaling  are evident in the low $T$ region, but these are only observable 
because we have perfect data (no statistical  errors). In real  
(i.e.\ noisy)  data such  small effects could easily be obscured by the 
noise, and might lead one to suppose that the correct value of $\theta$ 
were close to $-0.36$.

This tell us that the relation  $\xi \sim T^{-\nu}$, where
$\nu = -1/\theta$, is only  valid for rather small $T$. Corrections to
this form are important over most of the regime presented in Figure 1.
A striking  confirmation of  this is provided  by Figure 2,  where the
same data are plotted  against $L/\xi_{exact}$, where $\xi_{exact}$ is
the exact  correlation length for  each temperature. From Eq.\  (2) we
can identify this  length scale as $\xi_{exact} =  -1/\ln a$. The data
in Fig.\ 2 collapse almost perfectly for all sizes $L \ge 40$.

The lesson here  is that it is not FSS itself  which is breaking down,
but  the  use of  the  relation $\xi  \sim  T^{-\nu}$  over the  whole
temperature range  explored. We  suspect that similar  problems affect
the interpretation of the data of Kawashima et al.\ \cite{Kawashima92}.

\noindent\underline{\bf (ii) $T=0$, $h>0$}
For  non-zero  magnetic  field,  the  1-d  problem  cannot  be  solved
analytically in  closed form  for general system  size, $L$.  One can,
however,   determine   the   relevant   correlation   length,   $\xi$,
numerically. Here $\xi$ is to  be interpreted as the length scale over
which the  ground state for  $h>0$ locally resembles the  $h=0$ ground
state. We define it as the  average `domain length', where a domain is
a cluster of spins completely aligned with one of the two $h=0$ ground
states.   This  definition   corresponds  to   the   scaling  argument
$L^{d/2}h^{1/\delta}$  in Eq.\  (\ref{m}), i.e.\  this  argument $\sim
(L/\xi)^{d/2}$ with $\xi \propto h^{-2/d\delta}$.

\begin{figure}
\begin{center}
\epsfysize=7cm
\epsfbox{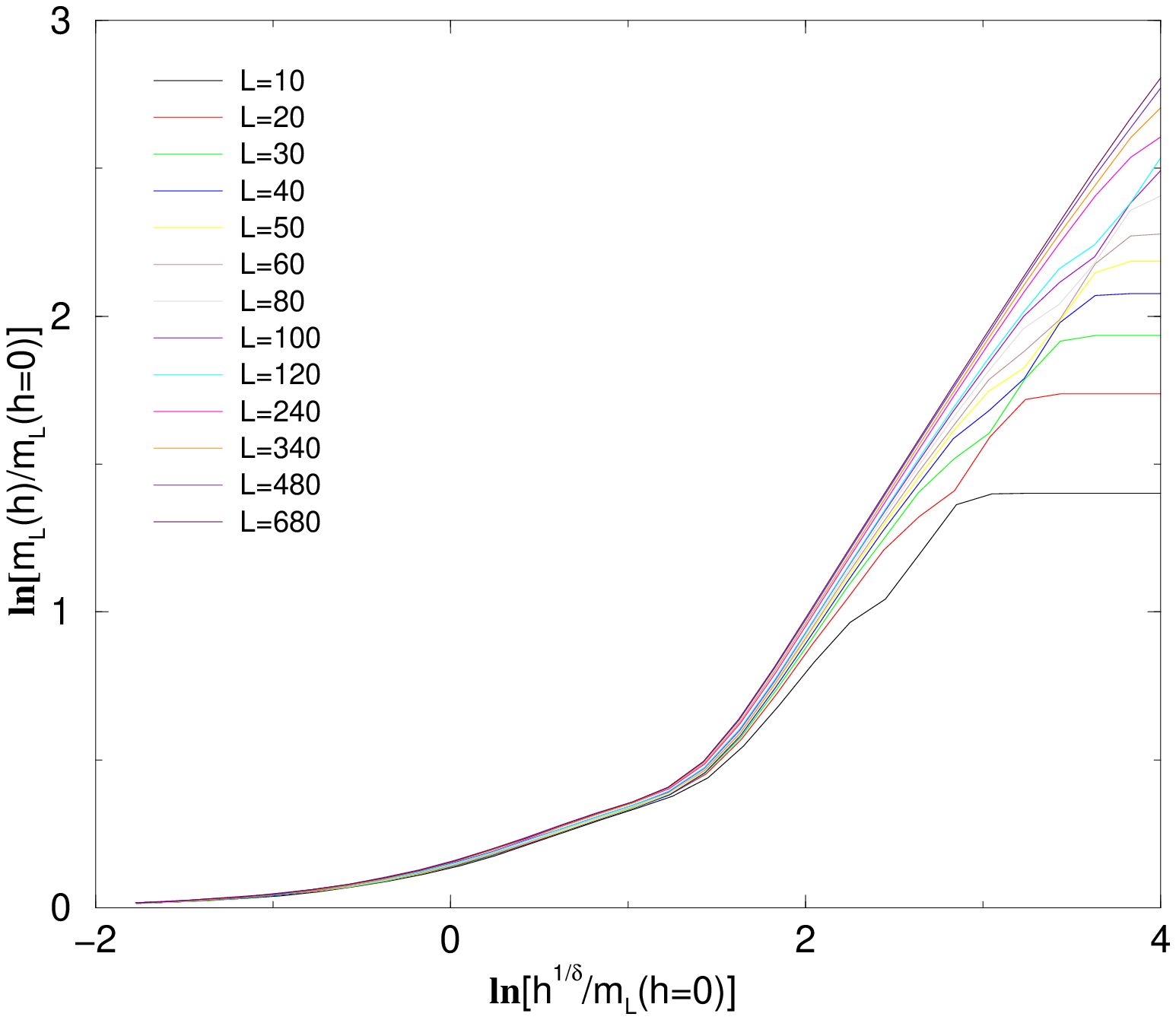}
\epsfysize=7cm
\epsfbox{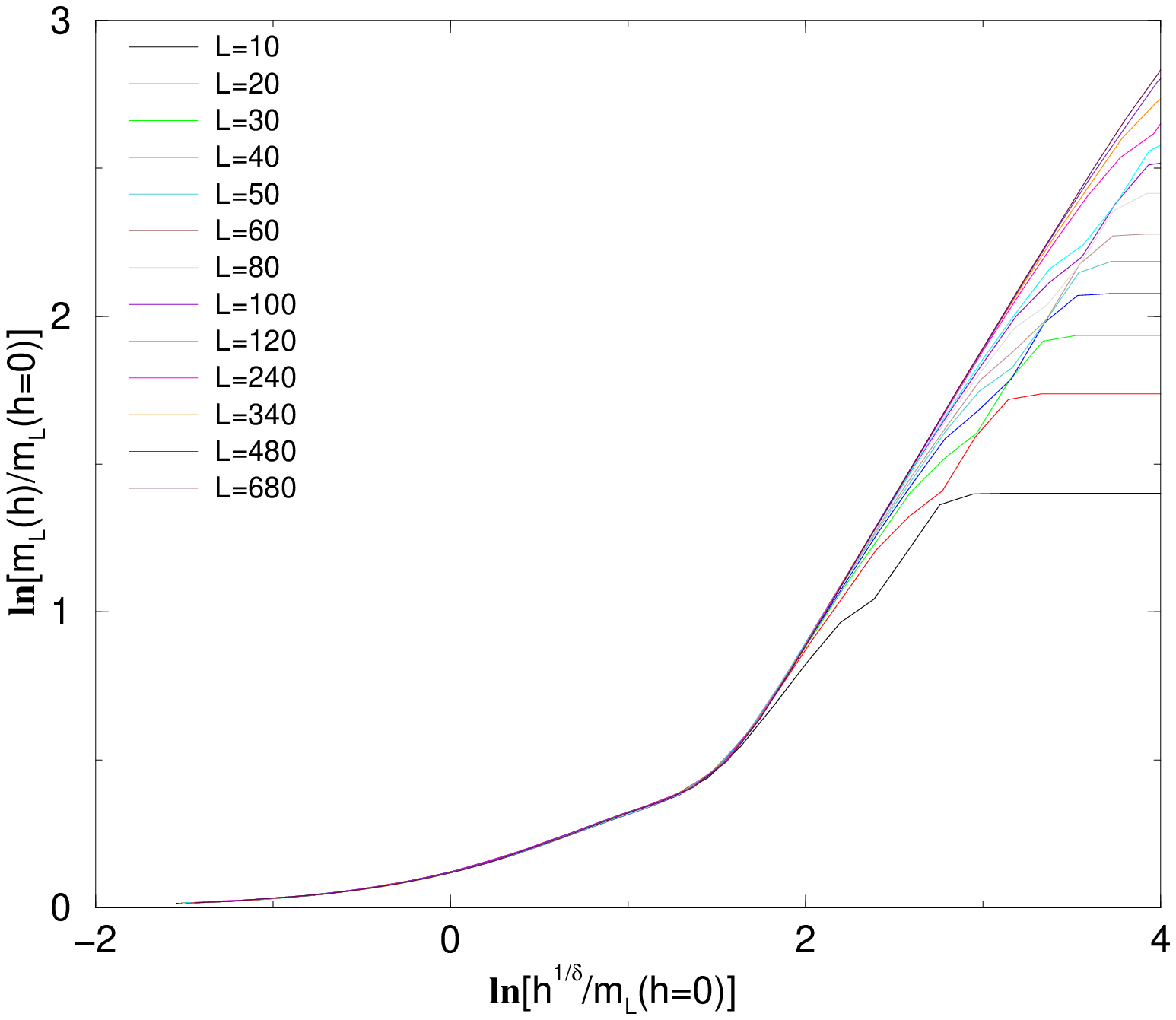}
\caption{Magnetisation versus field, scaled by the numerically
determined zero-field magnetization. \emph{Top}:
theoretical, $\delta=1.282$ and \emph{Bottom}: best collapse by eye, 
$\delta_{col}=1.37$ \label{fig:newscaling}}
\end{center}
\end{figure}
\begin{figure}
\begin{center}
\epsfysize=7cm
\epsfbox{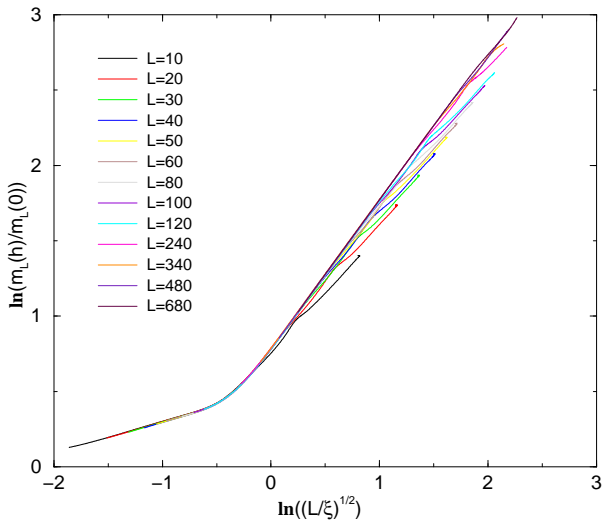}
\caption{Magnetization scaling plot using the numerically 
determined correlation length\label{fig:magcorlen}}
\end{center}
\end{figure}

Data for system sizes up to $L=680$ are displayed in Fig.\ 3, where each
data point  represents an average of $50\,000$  samples.  As explained
above, we choose  $\alpha = 6.042$ in (\ref{P}),  corresponding to the
value $\delta  = 1.282$ predicted by  the scaling theory  in $d=2$ for
$\theta = -0.282$. The plots are double-logarithmic plots based on the
asymptotic scaling  form (\ref{m}) with $d=1$.  It  is found, however,
that  if (\ref{m})  is used  as  it stands,  i.e.\ $L^{1/2}m_L(h)$  is
plotted against $L^{1/2}h^{1/\delta}$, it is difficult to get a perfect
data  collapse at  small  $h$, for  any  value of  $\delta$.  This  is
because  the magnetisation  per  spin  in zero  field  has a  binomial
distribution, and  its average over samples, $m_L(0)$,  is not exactly
proportional  to $L^{-1/2}$  for small  $L$ (for  $L \to  \infty$, the
binomial  distribution  approaches  the  normal distribution  and  the
$L^{-1/2}$ dependence  becomes asymptotically exact).   An alternative
FSS form  can, however,  be obtained by  simply replacing  the factors
$L^{1/2}$  in  (\ref{m}) by $1/m_L(0)$. This has been done in Fig.\ 3. 
In this way, the ordinate is identically zero at $h=0$ and data
collapse at small $h$ is greatly improved.

In the upper plot in Fig.\ 3, the exact value $\delta = 1.282$ is used,
while the lower plot shows the `best fit by eye' obtained with $\delta
\approx 1.37$.  The curves saturate at  large $h$, when  all spins are
aligned with  the field. The  scaling region extends to,  and somewhat
beyond, the  visible `elbow' in  the data. It  is a remarkable  that a
very good  fit by  eye (lower  plot) is obtained  for the  {\em wrong}
value,  $\delta =  1.37$, of  the  scaling exponent,  while the  exact
value, $\delta = 1.282$, is visibly worse (upper plot). 

This paradox, that  the best collapse appears to  be obtained with the
{\em wrong}  exponent value,  is resolved in  a similar manner  to the
$h=0$  case.   In  Fig.\ 4 we show the  same  data  replotted  against
$\ln[(L/\xi)^{1/2}]$,  where $\xi$ is the  average domain  length for  
a given $h$, as  defined above.  It is calculated numerically using 
transfer matrix methods and  taking systems long enough that the mean 
domain length converges (which requires longer systems at smaller $h$).

The collapse is very good, up to and a little beyond the elbow, as 
expected. The disappearance, in the new scaling variable, of the plateau 
from Fig.\ 3 is a consequence of the saturation of $\xi$, at the value 2, 
in this regime, i.e.\ the plateau gets compressed to a single point in 
these variables. As in the zero field case, the failure of the data to 
collapse with the correct scaling exponent, when naive FSS is employed, 
indicates an inappropriate choice of scaling variable rather than a 
failure of FSS itself, i.e.\ the scaling variable $L^{1/2}h^{1/\delta}$ 
is only useful at very small values of $h$.

Finally  we  have  studied  the  `domain wall' and  `droplet' energies
directly  for the  model  with bond  distribution (\ref{P}),  choosing
$\alpha = 2.546$, corresponding to  $\theta = -0.282$.  The system has
$L$ bonds and  free boundaries. The domain wall  energy, $E$, is given
by the  magnitude of the weakest  bond in the  system.  Averaging over
$10^5$ samples, with  $10 \le L \le 320$, and  plotting $\ln \langle E
\rangle$  against $\ln  L$ gives  the exponent  $\theta  = -0.297(2)$,
which is again different from  the exact value for this model, $\theta
=  -0.282$, but the  difference is  much smaller  than in  the studies
described earlier  in this paper. There  is also a  small curvature in
the  data: Extracting  the exponent  from  the last  two data  points,
$L=160$ and 320, gives $\theta = -0.288$, even closer to the exact 
value. 

One can  create a droplet excitation  by fixing the  boundary spins in
the  ground-state configuration  and reversing  the central  spin. The
droplet that forms around the  reversed spin is bounded by the weakest
bonds to its left and right. Computing the droplet energy numerically,
averaging over  $10^5$ samples  for $10 \le  L \le 320$,  and plotting
$\ln \langle E_{\rm droplet} \rangle$  against $\ln L$ as before gives
$\theta  = -0.303(3)$.   Using only  the  last two  data points  gives
$\theta  = -0.290$.  The  discrepancy between  the values  of $\theta$
obtained from  domain-wall and  droplet energies is  not statistically
significant, in  contrast to what is observed  in $d=2$ \cite{Ritort}.
In the  latter case, the  boundary of a  droplet forms a  closed loop,
which  may tend  to raise  its energy  due to  an  effective repulsion
between different parts of the interface, leading to large corrections
to scaling \cite{HartMoore}. In  $d=1$ the `interface' consists of two
isolated points and this effect is absent.

In summary, we have demonstrated by exact calculations in $d=1$ that a
naive use of FSS, in which the asymptotic form of the scaling variable
is employed, can lead to erroneous estimates of the scaling exponents,
while FSS itself works rather well. The `rogue' values of $\theta$ and
$\delta$ extracted form  these studies differ from the  true values in
the same  sense as  the corresponding exponents  in $d=2$  differ from
those obtained  using domain-wall estimates for $\theta$.  We conclude
that domain-wall  studies provide  the most reliable  determination of
the exponent $\theta$ in spin glasses.

\end{document}